\documentclass[conference]{IEEEtran}
\IEEEoverridecommandlockouts
\usepackage{cite}
\usepackage{amsmath,amssymb,amsfonts}
\usepackage{algorithmic}
\usepackage{algorithm2e}
\RestyleAlgo{ruled}
\usepackage{graphicx}
\usepackage[caption=false,font=footnotesize]{subfig}
\usepackage{textcomp}
\usepackage{xcolor}
\def\BibTeX{{\rm B\kern-.05em{\sc i\kern-.025em b}\kern-.08em
    T\kern-.1667em\lower.7ex\hbox{E}\kern-.125emX}}

\begin{document}

\title{The Wyner Variational Autoencoder for Unsupervised Multi-Layer Wireless Fingerprinting}

\author{
\IEEEauthorblockN{Teng-Hui Huang\IEEEauthorrefmark{1}, Thilini Dahanayaka\IEEEauthorrefmark{2}, Kanchana Thilakarathna\IEEEauthorrefmark{3}, Philip H.W. Leong\IEEEauthorrefmark{4} and Hesham El Gamal\IEEEauthorrefmark{5}}
\IEEEauthorblockA{\IEEEauthorrefmark{1}\IEEEauthorrefmark{4}\IEEEauthorrefmark{5}%
School of Electrical and Information Engineering, University of Sydney, NSW, Australia 2006}
\IEEEauthorblockA{%
\IEEEauthorrefmark{2}\IEEEauthorrefmark{3}School of Computer Science, University of Sydney, NSW, Australia 2006
}
\IEEEauthorblockA{email:\{tenghui.huang, 
 thilini.dahanayaka, kanchana.thilakarathna, philip.leong, hesham.elgamal\}@sydney.edu.au}
}

\maketitle

\begin{abstract}
 Wireless fingerprinting refers to a device identification method leveraging hardware imperfections and wireless channel variations as signatures. Beyond physical layer characteristics, recent studies demonstrated that user behaviors could be identified through network traffic, e.g., packet length, without decryption of the payload. Inspired by these results, we propose a multi-layer fingerprinting framework that jointly considers the multi-layer signatures for improved identification performance. In contrast to previous works, by leveraging the recent multi-view machine learning paradigm, i.e., data with multiple forms, our method can cluster the device information shared among the multi-layer features without supervision. Our information-theoretic approach can be extended to supervised and semi-supervised settings with straightforward derivations. In solving the formulated problem, we obtain a tight surrogate bound using variational inference for efficient optimization. In extracting the shared device information, we develop an algorithm based on the Wyner common information method, enjoying reduced computation complexity as compared to existing approaches. The algorithm can be applied to data distributions belonging to the exponential family class. Empirically, we evaluate the algorithm in a synthetic dataset with real-world video traffic and simulated physical layer characteristics. Our empirical results show that the proposed method outperforms the state-of-the-art baselines in both supervised and unsupervised settings.
\end{abstract}

\begin{IEEEkeywords}
Cross layer design, Wireless fingerprinting, Supervised learning, Unsupervised learning, Deep learning.
\end{IEEEkeywords}

\section{Introduction}
Wireless Fingerprinting refers to a device identification method that can uniquely identify the transmitter. Usually, this involves leveraging the effects of imperfections in the electronic devices used to construct the transmitter circuit which imparts measurable features in the physical layer, e.g., carrier frequency offsets, inphase/quadrature imbalance, out of band energy, etc. Hardware-specific techniques have been further extended to features beyond hardware imperfections, e.g., in the physical layer, the response of the medium along the transmit-receive path provides location-specific, frequency selective information~\cite{xiao2007fingerprints}. Recent works have also demonstrated that applying modern machine learning techniques can effectively discriminate subtle differences in characteristics and hence achieve improved device identification accuracy over conventional approaches~\cite{yu2019robust,riyaz2018deep,merchant2018deep}.

In the quest to improve wireless fingerprinting, several works have proposed combining physical and higher-layer features to defend against security attacks~\cite{moreira2018cross,sankhe2019no,wang2019xlf,shen2020fine,madarasingha2022videotrain++,li2018deep,dahanayaka2020understanding,8543573}. Such multi-layer wireless fingerprinting techniques significantly increase the difficulty of spoofing for an adversary and enhance identification accuracy to prevent privacy leakage. Nonetheless, to the authors' knowledge, there is a need for theoretically sound, computationally efficient approaches in integrating multi-layer features. Most prior research has either adopted heuristic objectives, relied on specific technologies and protocols, or has limited scalability as the number of available features increases~\cite{gu2018bf,robyns2017physical}. Furthermore, most machine learning-based approaches require labeled training samples, which can be prohibitively expensive to obtain in practice~\cite{jian2020deep}. Instead of relying on simulation-based data that simplifies the time-varying nature of wireless communications~\cite{al2020exposing}, we aim to provide a theoretic-founded unsupervised learning framework for multi-layer wireless fingerprinting. 


One of the challenges in multi-layer wireless fingerprinting is the extraction of the common information shared among the multi-layer signatures. This goal is closely related to multi-view learning, where data in multiple forms come in pairs, sharing a common randomness across the multi-view observations~\cite{2019Mvsurvey,8715409}. 


Recently, there has been a notable body of work adopting information-theoretic formulations for multi-view learning. The aim is to characterize the complexity-performance trade-off and develop efficient algorithms based on the derived insights~\cite{9154315,e22020213,9965818,8986754,wang2019deep,Xu2014LMMVIB,wan2021multi}. Among these, significant contributions have been made in supervised settings. However, fewer results have been reported for the unsupervised counterpart. For information-theoretic unsupervised multi-view learning, the Wyner's common information framework focuses on characterizing the common randomness from two correlated random variables~\cite{wyner1975common}. The framework has been applied to two correlated multi-view observations without labels~\cite{disGrad21}. Beyond two correlated sources, the characterization is further extended to Gaussian random variables with an arbitrary number of views and random vector settings~\cite{5766249,wynerCont16,relaxWynerInfo22}. For more general cases, variants of Wyner's formulation are introduced in literature where a computational challenge is identified due to the non-convex feasible set of the formulated optimization problem~\cite{kumar2014exact,disGrad21}. Nonetheless, in addressing the challenges, previous works either resort to heuristics methods~\cite{Xu2014LMMVIB}, are limited to special cases~\cite{relaxWynerInfo22} or provide fewer insights for large-scale cases~\cite{disGrad21}.

In contrast to previous works, we formulate the multi-layer wireless fingerprinting into a multi-view learning framework where each layer-feature mapped to a source of view observations. This allows for improved device identification performance with more sources layer-feature. Moreover, with multi-layer features the proposed framework can identify devices without supervision, enabled by extracting the shared information among the multi-layer features. In extracting the shared information, we adopt an information-theoretic approach that extends the framework to supervised and semi-supervised settings with straightforward derivations. We address the intractability of the formulated problem with variational inference techniques, arriving at a tight surrogate bound that can be optimized efficiently. Leveraging the Wyner common information, we develop an algorithm that can extract the shared device information whose computation complexity scales linearly with respect to the number of multi-layer features. The algorithm applies to data statistics that can be modelled as any member of the exponential family. Moreover, it is robust to the imbalance of the dimensionality of the multi-layer features. Empirically, we evaluate the proposed approach on a synthetic two-layer dataset consisting of real-world video traffic and simulated CSI data samples. Our reported results show that our method outperforms the state-of-the-art approaches in both supervised and unsupervised settings. Overall, we not only demonstrate the feasibility of improving device identification performance with multi-layer features, but also provide a method to achieve efficient multi-layer device identification.

\section{Problem Formulation}
We propose using multi-layer characteristics to improve the device identification performance. Define the $i^{th}$ multi-layer feature as $X_i$, and assume that there are $V$ available features. The goal is to identify the discrete device information $Z$ that generates the $V$ layer features. Note that since $Z$ is hidden, only $\{X_i\}_{i=1}^V$ are accessible. The task is modelled as an unsupervised multi-view clustering problem~\cite{2019Mvsurvey,8715409}, where $\{X_i\}_{i=1}^V$ represents the multi-view observations. 

Then to extract the view-shared common features $Z$ (the device information), we adopt the Wyner's common information framework~\cite{wyner1975common}, aiming to construct a stochastic common information encoder $P(Z|X^V), X^V:=(X_1,\cdots,X_V)$ from the following information-theoretic optimization problem:
\begin{IEEEeqnarray}{rCl}
\underset{P(Z|X^V)}{\text{minimize}}&\,&I(X^V;Z),\nonumber\\
\text{subject to}&\,&X_S\rightarrow Z\rightarrow X_{S^c},\,\forall S\subset[V],\IEEEeqnarraynumspace\IEEEyesnumber\label{eq:wyner_problem}
\end{IEEEeqnarray}
where $S$ denotes a partition of the multi-layer features $X^V$, i.e., $S\subset[V],S\cap S^c=\emptyset,S\cup S^c=[V]$. $[V]:=\{1,\cdots,V\}$; $X_S\rightarrow Z\rightarrow X_{S^c}$ represents a Markov chain relation (conditional independence) for all partitions $S\subset[V]$; $I(X^V;Z)$ the mutual information defined as:
\begin{IEEEeqnarray*}{rCl}
    I(X^V;Z):=\mathbb{E}_{X^V,Z}\left[\log\frac{P(X^V,Z)}{P(X^V)P(Z)}\right],
\end{IEEEeqnarray*}
where $\mathbb{E}[\cdot]$ is the expectation operator. $I(X_S;X_{S^c}|Z)$ is the conditional mutual information of the random variables $X_S,X_{S^c}$ conditioned on $Z$~\cite{cover1999elements}:
\begin{IEEEeqnarray*}{rCl}
    I(X_S;X_{S^c}|Z):=\mathbb{E}_{X^V,Z}\left[\log{\frac{P(X_S,X_{S^c}|Z)}{P(X_S|Z)P(X_{S^c}|Z)}}\right].
\end{IEEEeqnarray*}
Observe that in problem \eqref{eq:wyner_problem}, since the variable to optimize with is the conditional probability $P(Z|X^V)$, the dimensions of the variable scales exponentially $\mathcal{O}(|X|^V)$ with respect to the number of the multi-layer features $V$~\cite{relaxWynerInfo22}. To avoid the ``curse of dimensionality''~\cite{donoho2000high,10.1093/imamat/24.1.59,4766926}, we focus on a relaxed version of \eqref{eq:wyner_problem}, where $Z$ is restricted to be discrete:
\begin{IEEEeqnarray}{rCl}
\underset{\theta\in\Theta}{\text{minimize}}&\,&H(Z),\nonumber\\
\text{subject to}&\,& D_{KL}[P(X^V)\parallel P_\theta(X^V)]\leq \eta\IEEEeqnarraynumspace\IEEEyesnumber\label{eq:relaxed_wyner_problem}
\end{IEEEeqnarray}
where $\eta>0$; $H(Z)$ the Shannon entropy of $Z$ and the Kullback-Leibler (KL) divergence between two measures $\mu,\nu$ is denoted as $D_{KL}[\mu\parallel\nu]$:
\begin{IEEEeqnarray}{rCl}
    D_{KL}[\mu\parallel\nu]:=\mathbb{E}_{\mu}\left[\log{\frac{\mu}{\nu}}\right],
\end{IEEEeqnarray}
with $\mu,\nu$ defined over a proper support~\cite{cover1999elements}; Compared to Wyner's formulation, \eqref{eq:relaxed_wyner_problem} more relaxed in two ways. First, it allows an approximation error, measured by the KL divergence between the data distribution $P(X^V)$ to the parameterized distribution $P_\theta(X^V)$. Here, the parameter space is defined as $\Theta =\{\theta|P_\theta(Z)\in\Omega_Z,P_\theta(X_i|Z)\in\Omega_{i},\forall i\in[V]\}$ with $\Omega_Z$ denotes the probability simplex and $\Omega_i$ the compound probability simplex for the $i^{th}$ feature. Second, due to discrete $Z$, the conditional entropy $H(Z|X^V)\geq0$ so that $I(X^V;Z)=H(Z)-H(Z|X^V)\leq H(Z)$. Different from \eqref{eq:wyner_problem}, the parameters to optimize with in the relaxed problem \eqref{eq:relaxed_wyner_problem} are the marginal and conditional probabilities $P_\theta(Z)$ and $\{P_\theta(X_i|Z)\}_{i=1}^V)$. Additionally, the common information encoder is computed through marginalization of the probabilities:
\begin{IEEEeqnarray}{rCl}
    P_\theta(Z|X^V)=\frac{P_\theta(Z)\prod_{i=1}^VP_\theta(X_i|Z)}{\sum_{z'\in\mathcal{Z}}P_\theta(z')\prod_{j=1}^VP_\theta(X_j|z')}.\label{eq:common_enc_marg}
\end{IEEEeqnarray}
For illustration, let us consider a two multi-layer features $V=2$ special case in the following. By optimizing the relaxed problem with a Lagrange multiplier, the unconstrained form of \eqref{eq:relaxed_wyner_problem} is given as follows:
\begin{IEEEeqnarray}{rCl}
    \mathcal{L}_\theta&:=&H(Z)+\gamma\left\{D_{KL}[P(X_1,X_2)\parallel P_\theta(X_1,X_2)]-\eta\right\},\IEEEeqnarraynumspace\IEEEyesnumber\label{eq:two_view_lag_case_relax}
\end{IEEEeqnarray}
where the scalar $\gamma>0$ is a multiplier. Then for discrete entropy it is well-known that $H(Z)\leq \log{|\mathcal{Z}|}$. Therefore, for a fixed cardinality $|\mathcal{Z}|$, minimizing the Lagrangian \eqref{eq:two_view_lag_case_relax} reduces to minimizing the KL divergence between the joint distribution of the multi-layer observations $P(X_1,X_2)$ and the parameterized counterpart $P_\theta(X_1,X_2)$. This insight can be generalized to an arbitrary number of $V$, which results in the formation \eqref{eq:relaxed_wyner_problem}. The difficulty in solving the problem \eqref{eq:relaxed_wyner_problem} is that the joint distribution of the multi-layer observations $P(X^V)$ is intractable~\cite{kingma2013auto}. While $P(X^V)$ can be estimated through counting the available samples in small-scale discrete settings, for large-scale cases, the complexity grows exponentially and hence is infeasible. To address the intractability, a tight surrogate upper bound of the KL divergence in \eqref{eq:relaxed_wyner_problem} can be derived through the variational inference~\cite{poole2019variational,8588399,blei2017variational,kingma2013auto}:
\begin{IEEEeqnarray}{rCl}
    &{}&D_{KL}\left[P(X^V)\parallel P_\theta(X^V)\right]\nonumber\\
    &\leq& -H(X^V)-\mathbb{E}_{X^V,Q_z}\left[\log{\frac{P_\theta(Z)\prod_{i=1}^VP_\theta(X_i|Z)}{Q_\theta(Z|X^V)}}\right].\IEEEeqnarraynumspace\IEEEyesnumber\label{eq:dkl_vi_ub}
\end{IEEEeqnarray}
This follows from the derivation:
\begin{IEEEeqnarray}{rCl}
    &{}&D_{KL}[P(X^V)\parallel P_\theta(X^V)]\nonumber\\
    &=&D_{KL}\left[P(X^V)\parallel\sum_{z\in\mathcal{Z}}P_\theta(z)\prod_{i=1}^VP_\theta(X_i|z)\right]\nonumber\\
    &=& -H(X^V)\nonumber\\
    &&-\mathbb{E}_{X^V}\left[\log{\sum_{z\in\mathcal{Z}}P_\theta(z)\prod_{i=1}^VP_\theta(X_i|z)\frac{Q_\theta(z|X^V)}{Q_\theta(z|X^V)}}\right]\nonumber\\
    &\leq&-H(X^V)-\mathbb{E}_{X^V,Q_z}\left[\log{\frac{P_\theta(Z)\prod_{i=1}^VP_\theta(X_i|Z)}{Q_\theta(Z|X^V)}}\right],\IEEEeqnarraynumspace\IEEEyesnumber\label{eq:vi_bound_derive}
\end{IEEEeqnarray}
where the last line of \eqref{eq:vi_bound_derive} follows by the Jensen's inequality~\cite{cover1999elements}. $Q_\theta(Z|X^V)$ is the variational encoder to be designed and the bound is tight when $Q_\theta(Z|X^V)=P_\theta(Z|X^V)$ (defined in \eqref{eq:common_enc_marg}).

The upper bound \eqref{eq:dkl_vi_ub} is a multi-view version of the variational autoencoder (VAE)~\cite{kingma2013auto}. We therefore name the proposed method as the \textbf{W}yner \textbf{V}ariational \textbf{A}uto\textbf{E}ncoder (W-VAE). But we contrast the major differences to the VAE here. First, the VAE did not consider the Wyner common information condition for multi-layer features, i.e., $P(X^V|Z)=\prod_{i=1}^VP(X_i|Z)$. Second, we obtain the variational distribution $Q_\theta$ through marginalization \eqref{eq:impl_softmax_q} and predict the cluster distribution directly while the VAE reparameterizes a Gaussian representation and requires additional parameters to map the representations to prediction or reconstruction. Lastly, because of the marginalization W-VAE can cluster multi-layer features without supervision, but cluster labels or additional clustering algorithms applied on the representation are required when using VAE for clustering~\cite{8957256,kingma2014semi}.

Minimizing the surrogate upper bound \eqref{eq:dkl_vi_ub} is equivalent to maximizing the likelihood function:
\begin{IEEEeqnarray}{rCl}
    \mathcal{R}_\theta:=\mathbb{E}_{X^V,Q_z}\left[\log{\frac{P_\theta(Z)\prod_{i=1}^VP_\theta(X_i|Z)}{Q_\theta(Z|X^V)}}\right].\label{eq:reward_wvae}
\end{IEEEeqnarray}
Base on the derivation, our overall objective is to maximizing the reward $\mathcal{R}_\theta$ through judiciously chosen parameterized distributions $P_\theta(Z),P_\theta(X_i|Z),\forall i\in[V]$, and obtain the common information encoder $P_\theta(Z|X^V)$ through marginalization \eqref{eq:common_enc_marg}. 

\section{Log-Likelihood Parameterization}\label{sec:llr_implement}
\begin{figure*}[t]
\centering
\includegraphics[width=0.7\textwidth]{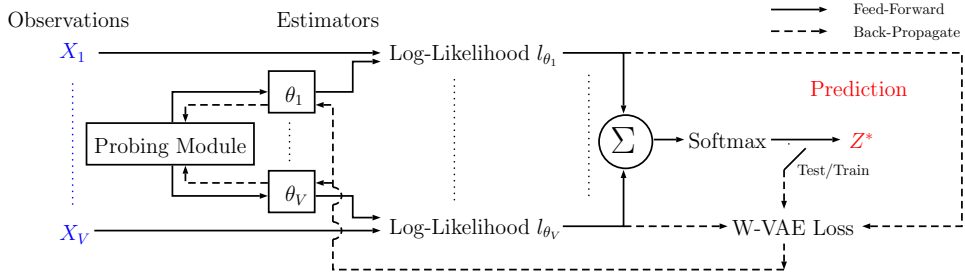}
\caption{The architecture of the Wyner variational autoencoder algorithm.}
\label{fig:block_diagram}
\end{figure*}
To simplify the problem, in the following we restrict the Wyner representation $Z$ to be uniformly discrete, i.e., the marginal distribution $P_\theta(Z)=1/|\mathcal{Z}|$ and denote it as $P(Z)$ for convenience of expression. Substituting these restrictions into \eqref{eq:reward_wvae}, the reward function can be rewritten as:
\begin{IEEEeqnarray}{rCl}
    \mathcal{R}'_\theta=\mathbb{E}_{X^V}\left[h_Q(X^V)\right]+\sum_{i=1}^V\mathbb{E}_{X_i,Q_z}\left[\log{P_\theta(X_i|Z)}\right],\label{eq:alg_reward_expand}
\end{IEEEeqnarray}
where in \eqref{eq:alg_reward_expand} the function $h_Q(X^V)$ is defined as $h_Q(x^v):=-\mathbb{E}_{Q_z}[\log{Q(Z|x^v)}]$. Operationally, given the multi-layer features $X^V$, \eqref{eq:alg_reward_expand} implies that the variational parameters corresponding to the individual layer-feature should be optimized through the maximum log-likelihood principle whereas the joint common information encoder is optimized from the maximum conditional entropy criterion~\cite{farnia2016minimax}.

The implementation of the objective function \eqref{eq:alg_reward_expand} can be divided into two parts. The first part is the variational encoder $Q_\theta(Z|X^V)$. From the derivation \eqref{eq:vi_bound_derive}, the surrogate upper bound is tight when $Q_\theta(Z|X^V)=P_\theta(Z|X^V)$, therefore it is desirable that the encoder satisfies the relation \eqref{eq:common_enc_marg}. The key observation is that it can be computed through a softmax function with the log-likelihoods as the inputs:
\begin{IEEEeqnarray}{rCl}
Q_\theta(Z|X^V)&=&\textit{Softmax}(\{\log{P_\theta(X_i|Z)}\}_{i=1}^V)\nonumber\\
&=&\frac{e^{\sum_{i=1}^V\log{P_\theta(X_i|Z)}}}{\sum_{z'\in\mathcal{Z}}e^{\sum_{i=1}^V\log{P_\theta(X_i|z')}}},\IEEEeqnarraynumspace\IEEEyesnumber\label{eq:impl_softmax_q}
\end{IEEEeqnarray}
where the last equality of \eqref{eq:impl_softmax_q} follows by the assumption that $P(Z)=1/|\mathcal{Z}|$. As for the conditional log-likelihoods $\log{P(X_i|Z)},\forall i\in[V]$, the implementation depends on the prior knowledge of the multi-layer observations. We list three practical and important classes of distributions as applications.
\subsection{Gaussian Mixture}
For multi-layer features that can be parameterized as a Gaussian distribution $X\sim\mathcal{N}(\boldsymbol{\mu}_Z,\Sigma_Z)$ when conditioned on the common information $Z$, e.g., physical layer (PHY) channel state information (CSI) and carrier phase offset, the log-likelihood is straightforward to derive. Suppose the conditional mean is $\boldsymbol{\mu}_Z$ and the conditional covariance matrix is $\Sigma_Z$, then the log-likelihood is given by:
\begin{IEEEeqnarray}{rCl}
    \log{P(X|Z)}&=&-\frac{1}{2}\log{(2\pi)^d|\Sigma_Z|}\nonumber\\
    &&-\frac{1}{2}(X-\boldsymbol{\mu}_Z)^T\Sigma_Z^{-1}(X-\boldsymbol{\mu}_Z),\IEEEeqnarraynumspace\IEEEyesnumber\label{eq:app_gmm_generic}
\end{IEEEeqnarray}
where $|\cdot|$ denotes the determinant operator. For convenience the multi-layer index (the subscript) of $X$ is omitted without loss of generality. Moreover, if the elements of the conditional Gaussian random vector $X$ are independent, then a simpler form of the log-likelihood can be derived:
\begin{IEEEeqnarray}{rCl}
    \log{P(X_{\perp}|Z)}&=&-\frac{d}{2}\log{2\pi}-\frac{1}{2}\sum_{j=1}^d\left[\log{\sigma^2_{j;Z}}\right.\nonumber\\
    &&\left.\vphantom{\log{\sigma}}+\frac{(x_{j}-\mu_{j;Z})^2}{\sigma^2_{j;Z}}\right],\IEEEeqnarraynumspace\IEEEyesnumber\label{eq:app_gmm_indep}
\end{IEEEeqnarray}
where $\mu_{j;Z},\sigma^2_{j;Z}$ are the mean and variance corresponding to the $j^{th}$ entry of the observation. Note that following the common practice where the mean and the log-variance $\nu_{j;Z}:=\log{\sigma^2_{j;Z}}$ are parameterized~\cite{kingma2013auto,alemi2016deep} to facilitate efficient optimization, we have another expression for \eqref{eq:app_gmm_indep}:
\begin{IEEEeqnarray}{rCl}
    \log{P(X_\perp|Z)}&=&-\frac{d}{2}\log{2\pi}-\frac{1}{2}\sum_{j=1}^d\left[\nu_{j;Z}\right.\nonumber\\
    &&\left.\vphantom{\nu}+(x_j-\mu_{j;Z})^2\exp\{-\nu_{j;Z}\}\right].\IEEEeqnarraynumspace\IEEEyesnumber\label{eq:app_gmm_parvar}
\end{IEEEeqnarray}

Using \eqref{eq:app_gmm_generic} and \eqref{eq:app_gmm_indep}, for multi-layer observations that are Gaussian mixture, the prediction of the common information representation can be obtained through the softmax function as shown in \eqref{eq:impl_softmax_q}. Examples that can be applied to this Gaussian mixture model include blind demodulation in additive white Gaussian noise (AWGN) channel~\cite{720245}, and unsupervised image clustering~\cite{e22020213}.
\subsection{Bernoulli Mixture}
For observations with binary outcomes, e.g., packet arrival/departure and (negative-) acknowledgement (ACK/NACK), one can parameterize them as conditional Bernoulli distributions. For simplicity of expression, we consider a $d$-dimensional binary vector $X$, where its elements are independent Bernoulli random variables. In this case, the log-likelihood has the following expression:
\begin{IEEEeqnarray}{rCl}
    \log{P(X|Z)}&=&\sum_{j=1}^d\left[\boldsymbol{1}\{x_j=1\}\log{\eta_j}\right.\nonumber\\
    &&\left.+\boldsymbol{1}\{x_j=0\}\log{(1-\eta_j)}\right],\IEEEeqnarraynumspace\IEEEyesnumber\label{eq:app_ber_origform}
\end{IEEEeqnarray}
where $\boldsymbol{1}\{\mathcal{A}\}$ denotes the indicator function of an argument $\mathcal{A}$, which outputs $1$ if $\mathcal{A}$ is true and $0$ otherwise; $\eta_j$ is the probability of positive outcome for the $j^{th}$ element of the observation $x_j$. In practice, a well-known trick to facilitate the estimation of the parameters is to parameterize the logarithm of the positive-to-negative probability ratio (the logit) instead. This results in the following equivalent expression of \eqref{eq:app_ber_origform}:
\begin{IEEEeqnarray}{rCl}
    \log{P(X|Z)}&=&\sum_{j=1}^dx_j\xi_j-\log{(1+e^{\xi_j})},\IEEEeqnarraynumspace\IEEEyesnumber\label{eq:app_ber_para}
\end{IEEEeqnarray}
where the logit is defined as $\xi_j:=\log{\eta_j/(1-\eta_j)}$; Note that the last term of \eqref{eq:app_ber_para} is the negative softplus function, where $\textit{softplus}(x):=\log{1+e^x}$. Similar to the Gaussian mixture case, once the Bernoulli distribution is parameterized, it can be substituted into \eqref{eq:impl_softmax_q} to compute the common information encoder.

\subsection{Generalization to the Exponential Family}
Following the previous discussions, the proposed method can be extended to the exponential family class of distributions, which include the previous examples as members. The family also includes important members in network quality of service (QoS) analysis, such as the Poisson and exponential distributions~\cite{vilaplana2014queuing,anastasi2000qos}. To include these classic statistic models as multi-layer features, we show in the following that our approach can be applied to the exponential family class of distributions. For convenience, we consider a vector observations $\boldsymbol{x}\in\mathbb{R}^d$, with independent components $x_j,j\in[d]$. For a single element $x_j$ and a given $z\in\mathcal{Z}$, we define the parameter vector $\boldsymbol{\eta}_{j;z}\in\mathbb{R}^{k_j}$, i.e., there are $k_j$ parameters corresponding to the $j^{th}$ entry of $\boldsymbol{x}$, conditioned on $z$. The resulting log-likelihood function is:
\begin{IEEEeqnarray}{rCl}
    \log{P(X|Z)}:=\sum_{j=1}^dh_j(x_j)+\boldsymbol{\eta}_{j;Z}^T\boldsymbol{T}_j(x_j)-A_j(\boldsymbol{\eta}_{j;Z}),\IEEEeqnarraynumspace\IEEEyesnumber\label{eq:app_gen_exp_family}
\end{IEEEeqnarray}
where $h(\cdot)$ is a normalization function independent of the parameters; $T(\cdot)$ denotes the sufficient statistic; $A(\cdot)$ the cumulant function, and the subscript indicates the observation entry. Note that the expression \eqref{eq:app_gen_exp_family} focuses on a single multi-layer feature, but it can be expanded to $\{X_i\}_{i=1}^V$ similarly.

\section{The Wyner Variational Autoencoder}
In the previous section, we provide details for the variables to optimize with for the problem \eqref{eq:relaxed_wyner_problem}. This consists of the parameterized conditional log-likelihoods for each multi-layer feature, along with the common information encoder computed from the softmax function \eqref{eq:impl_softmax_q}. Then we proceed to develop an algorithm to implement the loss (negative reward) function to update the parameterized distributions efficiently.

\subsection{Unsupervised Clustering}
Given the multi-layer features $\{X_i\}_{i=1}^V$ of $V$ signatures as the inputs, with a pre-determined cardinality of the common representation $Z$, the output is the soft-predictions of the conditional probability $P_\theta(Z|X^V)$, i.e., the distribution of the clusters $Z$ that a pair of multi-layer samples $x^V\in X^V$ belongs to. For each sample of the multi-layer observations $x^V\in X^V$, a set of $|\mathcal{Z}|$ parameterized conditional log-likelihoods $\{\log{P(X_i|Z)}\}_{i=1}^V$ are prepared through a probing module. By construction, the probing module should return the conditional log-likelihoods such that the parameters $\{\log{P(X_i|Z)}\}_{i=1}^V$ are independent across signatures, i.e., $X_i\perp X_j$ given a $z\in\mathcal{Z}$, for all $i\neq j\in[V]$. Within a single layer-feature $X_i$, a set of $|\mathcal{Z}|$ parameterized conditional log-likelihoods is prepared by the probing module for a given feature-specific observation $x_i$, then the resultant conditional log-likelihoods can be computed $\{\log{P(X_i|Z)}\}_{i=1}^V$ according to the associated equations (Gaussian, Bernoulli or other members of the exponential family). Finally, the conditional log-likelihoods are used to compute the common information encoder \eqref{eq:impl_softmax_q}.

We implement the algorithm as a deep neural network (DNN) where the pseudo-codes are described in Algorithm \ref{alg:wynerVAE} and the block diagram is shown in Fig. \ref{fig:block_diagram} for completeness. Due to the discrete restriction of $Z$, we use one-hot vectors $w_z,\forall z\in\mathcal{Z}$ (a vector where only a single element is $1$, and the other elements are all $0$s) as probing signals to prepare the parameterized conditional log-likelihoods, corresponding to each realization of the common representation $\forall z\in\mathcal{Z}$. The probing module $f:\mathcal{Z}\mapsto \{R^{|\mathcal{X}_i|}\}_{i=1}^V$ is implemented as a stack of neurons, and then the same probing module is connected to the individual parameterized conditional log-likelihoods independently. The design can be expressed as a composite function $\log{P(X_i|z)}:=g_i\circ f(w_z),\forall i\in[V],z\in\mathcal{Z}$. This satisfies the conditional independent condition $P_\theta(X^V|Z)=\prod_{i=1}^VP_\theta(X_i|Z)$ which in turns allows for linear growth rate of computation complexity $\mathcal{O}(V)$, defined as the parameters used w.r.t. the number of layer features $V$. We stress that the above description requires no knowledge of cluster labels, i.e., the ground-truth $Z^*$.
\begin{algorithm}
\caption{The Wyner Variational Autoencoder}\label{alg:wynerVAE}
\begin{algorithmic}
\STATE {\bfseries Input:}Multi-layer dataset $\mathcal{D}$ with $V$ sources of observations $(X_1,\cdots,X_V)$, cardinality of $\mathcal{Z}$
\STATE {\bfseries Output:}model weights $\theta^*$
\STATE {\bfseries Initialize:} Iteration counter $k=0$, weights $\theta_v\in\Theta_V$
\WHILE{$k \neq$ maximum number of epochs}
\STATE $z_i\gets \textit{onehot}(i)$ for each $i\in[|\mathcal{Z}|]$
\FOR{each $v\in[V]$}
\STATE $l_v\gets \textit{log-likelihood}_v(\{z_i\},x_v;\theta_{v})$, computed from a batch of $\{x_v\}_{v=1}^V\in\mathcal{D}$
\ENDFOR
\STATE $\hat{q}^{(k)}_\theta\gets \textit{softmax}(\sum_{v=1}^Vl_v)$
\STATE $\mathcal{R}_k\gets h_{\hat{q}^{(k)}_\theta}(x^V)+\sum_{z\in\mathcal{Z}}\sum_{v=1}^V\hat{q}^{(k)}_\theta\circ l_v$, \hfil {eq. \eqref{eq:alg_reward_expand}}
\STATE update $\{\theta_v^{k+1}\}_{v=1}^V\gets \textit{backpropagate}(-\mathcal{R}_k)$

\STATE $k\gets k+1$
\ENDWHILE
\end{algorithmic}
\end{algorithm}
\subsection{Supervised and Semi-Supervised Classifiers}
For the practical scenario where the dataset has a limited number of labels but not fully labeled, the proposed algorithm can leverage the available labels to improve the performance without changing the architecture. Consider the other extreme where the task is a fully supervised classification, i.e., each sample of the multi-layer features $x^V\in X^V$ has a cluster label $z^*\in\mathcal{Z}$. Then we can substitute the variational common information encoder with the ground-truth conditional probability $Q(Z^*|X^V)$ at the last equality of \eqref{eq:vi_bound_derive}, and arrive at the label-assisted loss upper bound:
\begin{IEEEeqnarray}{rCl}
    {}&&\mathcal{L}'_{\theta,V}\nonumber\\
    =&&-\mathcal{R}'_{\theta}+C_{\mathcal{Z}}\nonumber\\
    \leq&& \mathbb{E}_P\left\{D_{KL}[Q^*\parallel Q_\theta]-\sum_{i=1}^V\mathbb{E}_{Q^*}\left[\log{P_\theta(X_i|Z)}\right]\right\},\IEEEeqnarraynumspace\IEEEyesnumber\label{eq:semi_loss_ub}
\end{IEEEeqnarray}
where $Q^*:=Q(Z^*|X^V)$ denotes the ground-truth prediction for a given multi-layer feature sample, $Q_\theta:=Q_\theta(Z|X^V)$; $C_\mathcal{Z}$ is a constant independent of the parameters $\theta\in\Theta$ due to the assumption $P(Z)=1/|\mathcal{Z}|$ and the availability of the labels. The first term in the upper bound \eqref{eq:semi_loss_ub} is the standard cross-entropy (with $Q^*$ represents a one-hot vector), and the second term consists of the conditional log-likelihood functions that will be maximized for layer-feature specific estimators. The derivation follows by substituting the following into \eqref{eq:vi_bound_derive}:
\begin{IEEEeqnarray}{rCl}
    &{}&-\mathbb{E}_{P}\left[\log \sum_{z\in\mathcal{Z}}P_\theta(X^V|Z)P(Z)\frac{Q^*(Z|X^V)}{Q^*(Z|X^V)}\right]\nonumber\\
    &\leq&\mathbb{E}_{P}\left\{\mathbb{E}_{Q^*}\left[\log{\frac{Q^*(Z|X^V)}{P_\theta(X^V|Z)P(Z)}}\right]\right\}\nonumber\\
    &=&\mathbb{E}_{P}\left\{\mathbb{E}_{Q^*}\left[\log{\frac{Q^*(Z|X^V)}{Q_\theta(Z|X^V)}\frac{Q_\theta(Z|X^V)}{P(X^V|Z)P(Z)}}\right]\right\}\nonumber\\
    &\leq&\mathbb{E}_{P}\left\{D_{KL}[Q^*\parallel Q_\theta]+\mathbb{E}_{Q^*}\left[\log{\frac{Q^*(Z|X^V)}{P(Z)}}\right]\right.\nonumber\\
    &&\left.+\sum_{i=1}^V\mathbb{E}_{Q^*}\left[\log{P(X_i|Z)}\right]\right\},
    \IEEEeqnarraynumspace\IEEEyesnumber\label{eq:super_ub}
\end{IEEEeqnarray}
where the first inequality follows Jensen's inequality, and the second inequality is due to the non-negativity of the KL divergence. The bound is tight when the common information encoder, the variational distribution, and the labels information coincide, i.e., $P_\theta(Z|X^V)=Q_\theta(Z|X^V)=Q^*(Z|X^V)$, with $P_\theta(Z|X^V)$ computed from the marginalization \eqref{eq:common_enc_marg}. Note that the second term of the last line of \eqref{eq:super_ub} is a constant independent of the parameters $\theta$. Compared to the overall objective function for the unsupervised learning counterpart \eqref{eq:alg_reward_expand}, the maximum conditional entropy principle for the parameter updates of the common information encoder is replaced with the minimum cross-entropy loss with respect to the one-hot labels, but the same maximum log-likelihood criterion is imposed on the feature-specific estimators. 

Combining the objective functions in both unsupervised and supervised learning regimes, we obtained the semi-supervised variant of the W-VAE algorithm. Consider a semi-supervised scenario where the multi-feature dataset has $V$ sources. Among which, there are $N_u$ unlabeled samples and $N_l$ labeled samples ($N=N_u+N_l$ total number of samples). The empirical estimate of the objective function can be expressed as:
\begin{IEEEeqnarray}{rCl}
    &{}&N\mathcal{L}'_{\theta,V}\nonumber\\
    \approx&& -\sum_{v=1}^V\mathbb{E}_{Q^*_\theta}\left[\log{P(x_v|Z)}\right]\nonumber+\sum_{j\in\mathcal{N}_l}Q^*_j\log{\frac{Q^*_j}{Q_{\theta,j}}Q_{\theta,j}}\nonumber\\
    &&+\sum_{k\in\mathcal{N}_u}Q_{\theta,k}\log{Q_{\theta,k}}\nonumber\\
    =&&\sum_{m\in\mathcal{N}}\left\{-\sum_{v=1}^V\mathbb{E}_{Q^*_{\theta,m} }\left[\log{P_\theta(x_v|Z)}\right]+\mathbb{E}_{Q^*_{\theta,m} }\left[\log{Q_{\theta,m} }\right]\right\}\nonumber\\
    &&+\sum_{j\in\mathcal{N}_l}\mathbb{E}_{Q^*_{j}}\left[\log{\frac{Q^*_{j}}{Q_{\theta,j}}}\right],\IEEEeqnarraynumspace\IEEEyesnumber\label{eq:semi_loss}
\end{IEEEeqnarray}
where the notation $Q^*_\theta$ denotes the stack of predictions (each is a $\mathcal{Z}$ dimensional vector) substituted the columns that have labels with the ground-truth $Q^*$. Observe that in equation \eqref{eq:semi_loss}, if there is any label information available in the dataset, the cross-entropy (the last term) is used to include them into the objective function, then the parameters are updated from the backpropagation.

\subsection{Weighting the Multi-Layer Log-likelihoods}\label{sec:wvae_weighting}
In previous parts, an implicit assumption we made is that each multi-layer feature has approximately the same reliability (importance) for device identification. In practice, the reliability of the multi-layer features might vary significantly, e.g., imbalanced number of features for each layer-feature. In these cases, instead of treating the multi-layer features as equally reliable, a weighting of the reliability of sources could provide more robustness for the proposed algorithm to account for imbalance or corruption of certain layer-features. Motivated by this, we have the following variant of the log-likelihood terms in the W-VAE algorithm:
\begin{IEEEeqnarray}{rCl}
    \mathbb{E}_{Q}[\log{\prod_{i=1}^V}P^{V\alpha_i}(X_i|Z)]&=&V\sum_{i=1}^V\alpha_i\mathbb{E}_{Q}\left[\log{P_\theta(X_i|Z)}\right],\IEEEeqnarraynumspace\IEEEyesnumber\label{eq:eval_variant_alpha}
\end{IEEEeqnarray}
where $\forall i\in[V], \alpha_i>0,\sum_{i=1}^V\alpha_i=1$ denotes the weight for the $i^{th}$ multi-layer feature. Note that when $\alpha_i=1/V$, the log-likelihood terms reduce to the standard form \eqref{eq:alg_reward_expand}. Then, the weighting will be applied to each multi-layer specific log-likelihoods, resulting in a weighted softmax function:
\begin{IEEEeqnarray}{rCl}
    Q_\theta(Z|X^V;\alpha)=\textit{Softmax}\left(\sum_{i=1}^V\alpha_i\log{P_\theta(X_i|Z)}\right).
\end{IEEEeqnarray}
This variant is in resemblance of the general guideline in estimation theory, that is, when the prior probability for each multi-layer feature $\{\alpha_i\}_{i=1}^V$ is unknown, the \textit{Maximum (Log)-Likelihood} (ML) estimators are the optimal whereas the \textit{Maximum A Posteriori} (MAP) estimators give better performance from incorporating the knowledge of $\{\alpha_i\}_{i=1}^V$. It is also well-known that when the prior probability is uniform, $\{\alpha_i\}_{i=1}^V=1/V$, the two estimators coincide~\cite{kay1993fundamentals}. The remarks follow from the derivation of our theoretic formulation naturally and is another strength of our approach.

\begin{figure*}[!ht]
    \centering
    \includegraphics[width=0.70\textwidth]{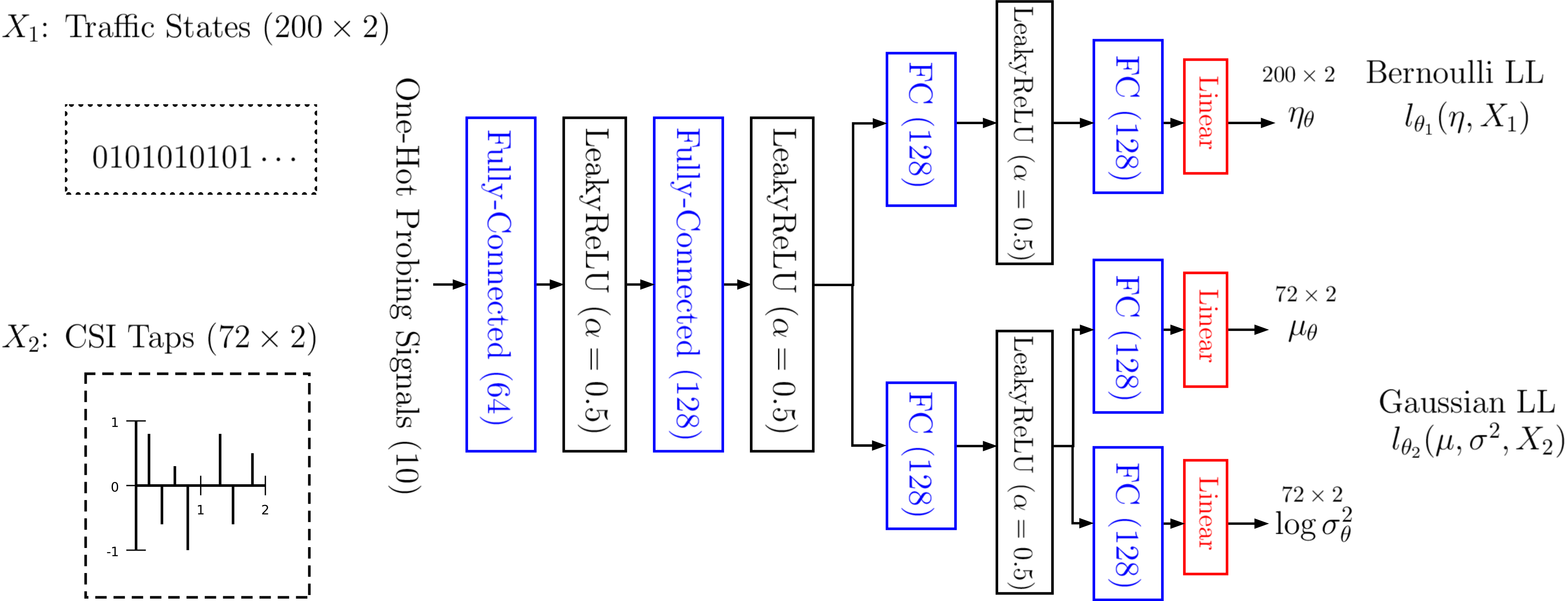}
    \caption{Network architecture of the implementation of Algorithm \ref{alg:wynerVAE}. The Gaussian and Bernoulli log-likelihoods (LL) follows \eqref{eq:app_gmm_parvar} and \eqref{eq:app_ber_para}.}
    \label{fig:tf_architecture}
\end{figure*}
\section{Evaluation}
We evaluate the proposed W-VAE algorithm on a synthetic multi-layer signature dataset. The synthetic dataset consists of a real-world video traffic dataset~\cite{li2022traffic} paired with a simulated channel state information (CSI). A sample of the video traffic dataset has $200$ binary sequences of the uplink and downlink packet lengths, and we pre-process them into traffic states ($0$: idle, $1$: non-zero packet lengths). As for the CSI dataset, a sample has $M=72$ complex values, each is computed from the standard least-square estimators with WLAN short preamble as the pilot signals ($72$ complex symbols) over a simulated Rayleigh fading channel:
\begin{equation}
    \boldsymbol{\hat{h}} =(\boldsymbol{X}^H\boldsymbol{C}^{-1}\boldsymbol{X})^{-1}\boldsymbol{X}^H\boldsymbol{C}^{-1}\boldsymbol{y},
\end{equation}
where $\boldsymbol{X}$ denotes the matrix form of the pilot signals (full rank); $\boldsymbol{y}$ the received signal and $\boldsymbol{\hat{h}}$ the channel estimate; $\boldsymbol{C}$ the noise covariance and we set $\boldsymbol{C}=\boldsymbol{I}$; The signal model is $\boldsymbol{y}:=\boldsymbol{X}\boldsymbol{h}'+\boldsymbol{w}$, $w_i\sim\mathcal{N}(0,\sigma_w^2),\forall i\in[M]$. The pilot signals $\boldsymbol{X}$ has $10$ dB signal-to-noise power ratio (SNR). For each class (the video ID), we generate $72\times 2$ standard normal distribution samples, reshaped into $72$ complex values as the mean vector $\boldsymbol{h}$ of the CSI. Then, to account for the time-varying natural of wireless channel, we manually add Gaussian noise with a configurable variance $\sigma_h^2$. In other words, $\boldsymbol{h}':=\boldsymbol{h}+\boldsymbol{\varepsilon},\varepsilon_i\sim\mathcal{N}(0,\sigma_h^2),\forall i\in[M]$, and higher CSI variance $\sigma_h^2$ will degrade the classification/clustering performance. To control the CSI variation, we introduce the metric: CSI Perturbation-to-noise ratio (PNR), $\text{PNR}={\sigma_h^2}/{\sigma_w^2}$ in the following experiments.

We combine the two datasets with the video traffic sequences as the first multi-layer feature and the CSI as the second one. There are $10$ videos traffic sequences collected from YouTube as detailed in~\cite[Sec 3.2]{li2022traffic}. 
After pairing each video sample with a simulated CSI, there are $2557$ training samples and $638$ testing samples with each set uniformly distributed across the $10$ classes.

The two-layer dataset matches the following device identification task with the two-layer features. Consider a wireless network where there are $10$ devices streaming videos from a platform over the same router. For simplicity, we assume that all the devices are using the same streaming platform, and the same wireless technology. The router serves all the devices, and the proposed algorithm is implemented at the router's side, with access to the physical CSI of the devices and the network layer traffic states. Through monitoring the streaming traffic states and the wireless CSI, the proposed algorithm's goal is to identify the device that could potentially stream malicious contents from the accessible features. The router cannot examine the content directly since the network traffic is typically encrypted and no decipher is available at the router, but the packet lengths (traffic states) are accessible. Under this setup, the task is equivalent to a supervised classification, or an unsupervised multi-view clustering problem depending on the availability of device labels. As for the feasibility of the physical layer CSI, it can be estimated from feedback. The feedback signals can be the uplink transmission, or the control channel signals in cellular networks, carrying the normal payload or ACK/NACK in the multiple access (MAC) layer.

 We implement Algorithm \ref{alg:wynerVAE} on Tensorflow2. The architecture follows the block diagram in Fig. \ref{fig:block_diagram} whose network design details are provided in Fig. \ref{fig:tf_architecture} for completeness. The probing module is implemented as a stack of fully connected layers with leaky rectified linear unit (Leaky ReLU) activation to avoid vanishing gradients~\cite{xu2015empirical}. The common information $Z$ is enforced to be a discrete random variable serving the cluster labels. Given a pre-determined number of clusters $|\mathcal{Z}|$, the hidden layer output of the networks is used to compute the conditional log-likelihoods and obtain the prediction through \eqref{eq:impl_softmax_q}. The conditional log-likelihood for each layer-feature is implemented as separate fully connected layers with linear activation functions. For the traffic states $X_1$, we model them as a Bernoulli mixture where the re-parameterized probability of positive outcomes $\eta_{\theta}$ is obtained from the hidden layer output and used to calculate the log-likelihood $l_{\theta_1}$ \eqref{eq:app_ber_para} with $X_1$. As for the CSI taps, we model them as Gaussian mixture where the re-parameterized means $\mu_\theta$ and log-variances $\log{\sigma_\theta^2}$ are obtained from the hidden layer outputs and used to calculate the log-likelihood $l_{\theta_2}$ \eqref{eq:app_gmm_generic} with $X_2$.

\begin{figure*}[!t]
    \centering
    \subfloat[Versus Baseline using traffic only]{%
        \centering
        \includegraphics[width=0.4\textwidth]{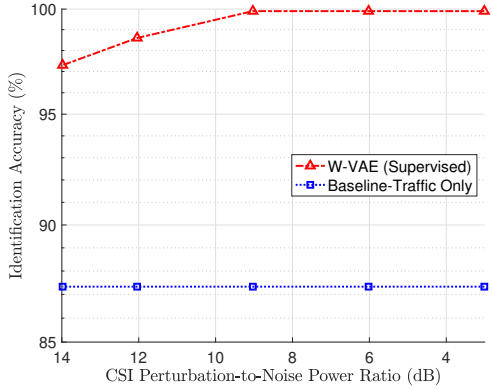}
        \label{subfig:su_traffic_only}
    }
    \hfil
    \subfloat[Versus Baseline with both traffic and CSI]{%
        \centering
        \includegraphics[width=0.4\textwidth]{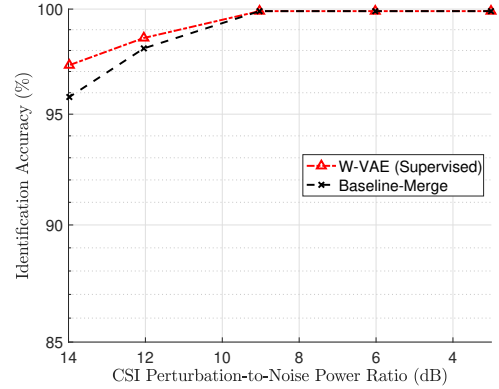}
        \label{subfig:su_both}
    }
    \caption{Identification accuracy versus Perturbation-to-Noise Power Ratio (PNR) of the CSI (Supervised). The Baseline follows \cite{li2022traffic}. The W-VAE (Supervised) uses both traffic (higher layer) and CSI (PHY layer) features.}
    \label{fig:su_perturb_acc}
\end{figure*}

\subsection{Supervised Device Identification}
We first evaluate the proposed approach in a supervised classification task. For simplicity, we assume knowledge of the optimal number of clusters $Z=10$ and a discussion for relaxing this knowledge is deferred to Section \ref{sec:relax_label}. The total number of parameters is approximately $9.78\times 10^{4}$. The loss function for the supervised variant of the W-VAE follows \eqref{eq:super_ub} and we denote it as \textit{W-VAE (Supervised)}. We compared the \textit{W-VAE (Supervised)} with a baseline approach~\cite{li2022traffic}. This compared method implemented a deep neural network (DNN)-based classifier and empirically demonstrated the state-of-the-art classification accuracy with the video traffic dataset. This baseline only uses traffic states as the inputs and is denoted as \textit{Traffic Only} baseline. The DNN consists of two fully connected layers with dropout~\cite{srivastava2014dropout}, the details are referred to \cite[Fig. 5]{li2022traffic}. We modify the number of neurons to $200\rightarrow128$, resulting 
 in approximately $1.07\times 10^{5}$ number of parameters.
 
 For each method in the the evaluated approaches, $25$ trials are performed. Each trial runs the training dataset for $200$ epochs, determined from cross-validation and the classification accuracy is computed offline from the testing dataset. For the W-VAE (supervised), $8$ mini-batch size is used with a fixed learning rate $10^{-3}$ for the standard ADAM optimizer~\cite{kingma2014adam}. As for the baseline, the configurations follow~\cite[Section 4.2.3]{li2022traffic}. We report the model with the maximum accuracy from $25$ trained ones.
 The results are shown in Fig. \ref{fig:su_perturb_acc}. In Fig. \ref{subfig:su_traffic_only} compared to the \textit{Traffic-Only} baseline, the W-VAE (Supervised) improves the classification accuracy significantly over the range of PNR$\in[3,14]$ dB. Moreover, when $\text{PNR}<9$ dB, the W-VAE (Supervised) achieves $>99\%$ accuracy. 
 
 For a fair comparison and to highlight the efficiency of W-VAE, we further merge the traffic states and the CSI estimates as the input. To keep the total number of parameters at the same order, the hidden layer neurons are configured to $150\rightarrow 150$ (around $1.06\times 10^{5}$ parameters). The modified baseline is denoted as \textit{Merge}. In Fig. \ref{subfig:su_both}, compared to the \textit{Merge} baseline, the W-VAE (Supervised) can also attain slightly better performance over the simulated range of PNR with the same order of the number of parameters used.
\begin{figure}[!t]
    \centering
    \includegraphics[width=2.9in]{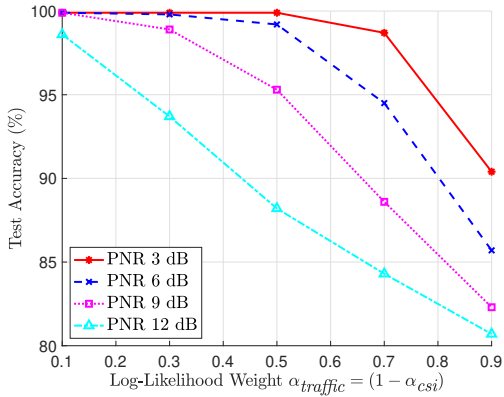}
    \caption{Test accuracy versus weighting ratio for W-VAE (Supervised).}
    \label{fig:su_ratio}
\end{figure}
In Fig. \ref{fig:su_perturb_acc}, the weighting for the W-VAE (Supervised) is $\alpha_{\text{traffic}}=0.1 (\alpha_{\text{csi}}=0.9)$. As detailed in Section \ref{sec:wvae_weighting}, the parameter search for $\alpha_{\text{traffic}}$ ($\alpha_{\text{csi}}$) depends on the knowledge of the prior distributions for the multi-layer features, and how to estimate the feature priors is out of the scope of this work. Here, we determine this hyperparameter empirically. We run five possible choices $\alpha_{\text{traffic}}\in[0.1,0.3,0.5,0.7,0.9]$ and select the best performing prior as the multi-layer feature prior probability. The detailed results over the same range of PNR in the first experiment are reported in Fig. \ref{fig:su_ratio}.

The two results demonstrate that the proposed method can successfully integrate the multi-layer features in an efficient and theoretic-founded fashion for improved performance.


\subsection{Unsupervised Device Identification}

Then we evaluate the proposed method in unsupervised clustering settings with the same dataset. In this case, the models are trained without access to the labels. For the W-VAE, the same model architecture (same number of parameters) is reused since only the loss (negative reward) function is changed to \eqref{eq:reward_wvae}. However, without label information the trained model's performance relies more on the initialization point. Therefore, the number of trails in this setting is set to $40$. As in the last experiment setup, we assume knowledge of the number of clusters of the dataset and set the number of training epochs to $200$. The best model is reported according to the lowest total training loss value the model achieved. We report the model with the lowest loss among the $40$ trained models. For testing performance, note that without supervision, the predicted clusters do not necessarily match the indices of the labels. Therefore, label matching is performed to obtain the testing accuracy. This can be done either using exhaustive search over all combination of label assignment which has $10!$ possibilities but no label is required or using a handful of labels for one-shot learning as in unsupervised clustering literature~\cite{xie2016unsupervised} which significantly reduces computation complexity. Since the synthetic dataset we adopted has labels, we follow the latter approach, but we stress that this label information is used exclusively for label matching purposes and is inaccessible during model training phase. 

We compare our method to two baselines. The first is a K-means-based method~\cite{kirchler2016tracked}. We use an off-the-shelf K-means implementation~\cite{scikit-learn}, with input features formed from cascading the $200\times 2$ bits traffic states and the $72\times 2$ real value CSI (real and imaginary number as two independent channels). The configurations of the hyperparameters are set to the default values as in~\cite[KMeans]{scikit-learn}. The evaluation of the testing phase performance follows the same label matching procedures as adopted in the W-VAE.

The second method is the contrastive fusion network (\textit{Conan})~\cite{9671851}, which is the state-of-the-art method in multi-view learning literature. This method has two view-specific representation encoders and contrastive metrics~\cite{tian2020contrastive} are adopted to fuse the two separate representations. We produce results for the second benchmark from the available prototype~\cite{9671851} with default parameters but modify the two-layer fully-connect layers of $128$ neurons for each view-specific encoder. The total number of parameters is $3.23\times 10^{5}$.

\begin{figure}[t]
    \centering
    \includegraphics[width=2.9in]{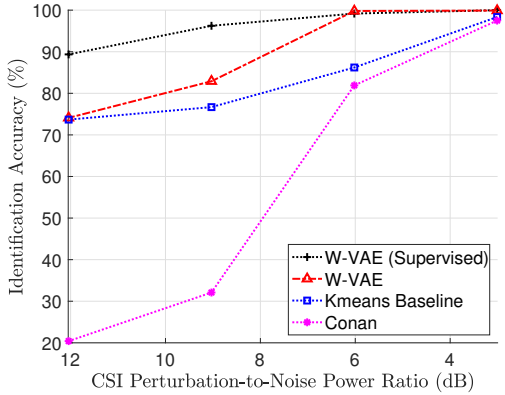}
    \caption{Identification accuracy versus PNR of the CSI (Unsupervised). Both methods use the two multi-layer features (traffic states and PHY CSI) without using the label information. The baseline follows~\cite{kirchler2016tracked}. }
    \label{fig:unsu_perturb_acc}
\end{figure}

\begin{figure}[t]
    \centering
    \includegraphics[width=2.9in]{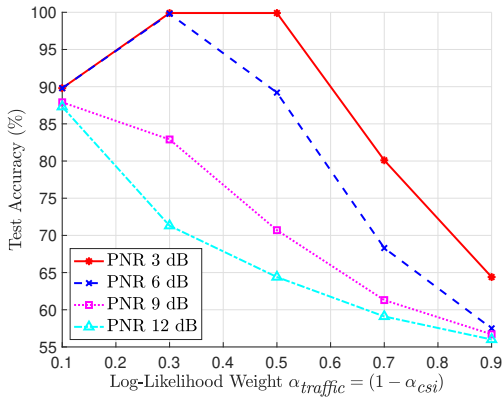}
    \caption{Test accuracy versus the weighting ratio for W-VAE.}
    \label{fig:unsupervised_ratio}
\end{figure}

The results are shown in Fig. \ref{fig:unsu_perturb_acc}, where the testing accuracy versus the range of PNR$\in[3,12]$ dB is reported. Over the range of PNR, the W-VAE outperforms the K-Means baseline with significantly higher clustering accuracy. In this experiment, we set the multi-layer weighting priors $\alpha_{\text{traffic}}=0.3$ ($\alpha_{\text{csi}}=0.7$), determined empirically in a separate experiment (detailed in Fig. \ref{fig:unsupervised_ratio}). Note that for the straightforward cascading of the multi-layer features, as the K-Means baseline and the case $\alpha_{\text{traffic}}=\alpha_{\text{csi}}=0.5$ in Fig. \ref{fig:unsupervised_ratio}, the clustering performance is sub-optimal, and the W-VAE demonstrates the benefit of a theoretic-founded and efficient approach to weight the multi-layer features for improved clustering performance. 


\subsection{Detecting the Optimal Number of Clusters}\label{sec:relax_label}
\begin{figure}[t]
    \centering
    \includegraphics[width=2.9in]{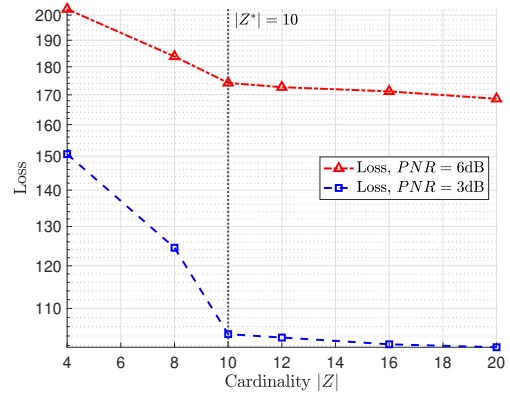}
    \caption{W-VAE loss value versus the the number of clusters $|\mathcal{Z}|$. The optimal $|\mathcal{Z}^*|=10$ is detectable by increasing $|\mathcal{Z}|$ from a small value. }
    \label{fig:latent_dim_detect}
\end{figure}
In this part, we relax the assumption of knowing the number of clusters of the dataset. This can be achieved through comparing the achieved loss value over a collection of trained models with different numbers of clusters. In practice, one can start with a small value of the cardinality of $Z$, e.g., $|\mathcal{Z}|=2$, train the model and record the loss value, increase the cardinality and repeat until reaching a sharp transition of the loss value versus $|\mathcal{Z}|$ as shown in Fig. \ref{fig:latent_dim_detect}. The detection of this transition can be done through a combination of change point detection algorithms, and standard binary or linear search solvers to identify the optimal cardinality of $Z$. Observe the results in Fig. \ref{fig:latent_dim_detect}, for two different settings of the CSI PNR, the optimal number of clusters all located at a sharp transition of the loss value versus $|\mathcal{Z}|$ and therefore can be detected. This demonstrates that the W-VAE can also be applied to dataset without a known number of clusters.

\section{Conclusion}
We propose a multi-layer wireless fingerprinting method leveraging signatures across layers which jointly improves the device identification performance. Adopting the multi-view machine learning paradigm allows for unsupervised clustering of the shared device information among multi-layer features. Our information-theoretic formulation can be extended to supervised and semi-supervised settings with straightforward derivations. In solving the intractability of the formulated problem, we adopt variational inference techniques leading to a tight surrogate bound. Then we propose extracting the shared device information through Wyner common information framework, leading to the development of the W-VAE algorithm for efficient multi-layer feature clustering with linear computation complexity as the number of layer features grows. The generic W-VAE algorithm can be parameterized as any member of the exponential family class of distributions and efficiently optimized with deep learning methods. The W-VAE is evaluated on a multi-layer dataset with network layer traffic and physical layer CSI. Our empirical results demonstrate that in both supervised and unsupervised scenarios, the W-VAE algorithm outperforms the state-of-the-art.

\bibliographystyle{IEEEtran}
\bibliography{references}

\begin{thebibliography}{10}
\providecommand{\url}[1]{#1}
\csname url@samestyle\endcsname
\providecommand{\newblock}{\relax}
\providecommand{\bibinfo}[2]{#2}
\providecommand{\BIBentrySTDinterwordspacing}{\spaceskip=0pt\relax}
\providecommand{\BIBentryALTinterwordstretchfactor}{4}
\providecommand{\BIBentryALTinterwordspacing}{\spaceskip=\fontdimen2\font plus
\BIBentryALTinterwordstretchfactor\fontdimen3\font minus
  \fontdimen4\font\relax}
\providecommand{\BIBforeignlanguage}[2]{{%
\expandafter\ifx\csname l@#1\endcsname\relax
\typeout{** WARNING: IEEEtran.bst: No hyphenation pattern has been}%
\typeout{** loaded for the language `#1'. Using the pattern for}%
\typeout{** the default language instead.}%
\else
\language=\csname l@#1\endcsname
\fi
#2}}
\providecommand{\BIBdecl}{\relax}
\BIBdecl

\bibitem{xiao2007fingerprints}
L.~Xiao, L.~Greenstein, N.~Mandayam, and W.~Trappe, ``Fingerprints in the
  ether: Using the physical layer for wireless authentication,'' in \emph{2007
  IEEE International conference on communications}.\hskip 1em plus 0.5em minus
  0.4em\relax IEEE, 2007, pp. 4646--4651.

\bibitem{yu2019robust}
J.~Yu, A.~Hu, G.~Li, and L.~Peng, ``A robust {RF} fingerprinting approach using
  multisampling convolutional neural network,'' \emph{IEEE internet of things
  journal}, vol.~6, no.~4, pp. 6786--6799, 2019.

\bibitem{riyaz2018deep}
S.~Riyaz, K.~Sankhe, S.~Ioannidis, and K.~Chowdhury, ``Deep learning
  convolutional neural networks for radio identification,'' \emph{IEEE
  Communications Magazine}, vol.~56, no.~9, pp. 146--152, 2018.

\bibitem{merchant2018deep}
K.~Merchant, S.~Revay, G.~Stantchev, and B.~Nousain, ``Deep learning for rf
  device fingerprinting in cognitive communication networks,'' \emph{IEEE
  journal of selected topics in signal processing}, vol.~12, no.~1, pp.
  160--167, 2018.

\bibitem{moreira2018cross}
C.~M. Moreira, G.~Kaddoum, and E.~Bou-Harb, ``Cross-layer authentication
  protocol design for ultra-dense 5{G} hetnets,'' in \emph{2018 IEEE
  International Conference on Communications (ICC)}.\hskip 1em plus 0.5em minus
  0.4em\relax IEEE, 2018, pp. 1--7.

\bibitem{sankhe2019no}
K.~Sankhe, M.~Belgiovine, F.~Zhou, L.~Angioloni, F.~Restuccia, S.~D’Oro,
  T.~Melodia, S.~Ioannidis, and K.~Chowdhury, ``No radio left behind: Radio
  fingerprinting through deep learning of physical-layer hardware
  impairments,'' \emph{IEEE Transactions on Cognitive Communications and
  Networking}, vol.~6, no.~1, pp. 165--178, 2019.

\bibitem{wang2019xlf}
A.~Wang, A.~Mohaisen, and S.~Chen, ``{XLF}: A cross-layer framework to secure
  the internet of things ({IoT}),'' in \emph{2019 IEEE 39th International
  Conference on Distributed Computing Systems (ICDCS)}.\hskip 1em plus 0.5em
  minus 0.4em\relax IEEE, 2019, pp. 1830--1839.

\bibitem{shen2020fine}
M.~Shen, Y.~Liu, L.~Zhu, X.~Du, and J.~Hu, ``Fine-grained webpage
  fingerprinting using only packet length information of encrypted traffic,''
  \emph{IEEE Transactions on Information Forensics and Security}, vol.~16, pp.
  2046--2059, 2020.

\bibitem{madarasingha2022videotrain++}
C.~Madarasingha, S.~R. Muramudalige, G.~Jourjon, A.~Jayasumana, and
  K.~Thilakarathna, ``{V}ideo{T}rain++: {GAN}-based adaptive framework for
  synthetic video traffic generation,'' \emph{Computer Networks}, vol. 206, p.
  108785, 2022.

\bibitem{li2018deep}
Y.~Li, Y.~Huang, R.~Xu, S.~Seneviratne, K.~Thilakarathna, A.~Cheng, D.~Webb,
  and G.~Jourjon, ``Deep content: Unveiling video streaming content from
  encrypted {WiFi} traffic,'' in \emph{2018 ieee 17th international symposium
  on network computing and applications (nca)}.\hskip 1em plus 0.5em minus
  0.4em\relax IEEE, 2018, pp. 1--8.

\bibitem{dahanayaka2020understanding}
T.~Dahanayaka, G.~Jourjon, and S.~Seneviratne, ``Understanding traffic
  fingerprinting {CNN}s,'' in \emph{2020 IEEE 45th Conference on Local Computer
  Networks (LCN)}.\hskip 1em plus 0.5em minus 0.4em\relax IEEE, 2020, pp.
  65--76.

\bibitem{8543573}
D.~Wang, B.~Bai, W.~Zhao, and Z.~Han, ``A survey of optimization approaches for
  wireless physical layer security,'' \emph{IEEE Communications Surveys \&
  Tutorials}, vol.~21, no.~2, pp. 1878--1911, 2019.

\bibitem{gu2018bf}
T.~Gu and P.~Mohapatra, ``{BF-IoT}: Securing the {IoT} networks via
  fingerprinting-based device authentication,'' in \emph{2018 IEEE 15Th
  international conference on mobile ad hoc and sensor systems (MASS)}.\hskip
  1em plus 0.5em minus 0.4em\relax IEEE, 2018, pp. 254--262.

\bibitem{robyns2017physical}
P.~Robyns, E.~Marin, W.~Lamotte, P.~Quax, D.~Singel{\'e}e, and B.~Preneel,
  ``Physical-layer fingerprinting of lora devices using supervised and
  zero-shot learning,'' in \emph{Proceedings of the 10th ACM Conference on
  Security and Privacy in Wireless and Mobile Networks}, 2017, pp. 58--63.

\bibitem{jian2020deep}
T.~Jian, B.~C. Rendon, E.~Ojuba, N.~Soltani, Z.~Wang, K.~Sankhe, A.~Gritsenko,
  J.~Dy, K.~Chowdhury, and S.~Ioannidis, ``Deep learning for rf fingerprinting:
  A massive experimental study,'' \emph{IEEE Internet of Things Magazine},
  vol.~3, no.~1, pp. 50--57, 2020.

\bibitem{al2020exposing}
A.~Al-Shawabka, F.~Restuccia, S.~D’Oro, T.~Jian, B.~C. Rendon, N.~Soltani,
  J.~Dy, S.~Ioannidis, K.~Chowdhury, and T.~Melodia, ``Exposing the
  fingerprint: Dissecting the impact of the wireless channel on radio
  fingerprinting,'' in \emph{IEEE INFOCOM 2020-IEEE Conference on Computer
  Communications}.\hskip 1em plus 0.5em minus 0.4em\relax IEEE, 2020, pp.
  646--655.

\bibitem{2019Mvsurvey}
Y.~Li, M.~Yang, and Z.~Zhang, ``A survey of multi-view representation
  learning,'' \emph{IEEE Transactions on Knowledge and Data Engineering},
  vol.~31, no.~10, pp. 1863--1883, 2019.

\bibitem{8715409}
W.~Guo, J.~Wang, and S.~Wang, ``Deep multimodal representation learning: A
  survey,'' \emph{IEEE Access}, vol.~7, pp. 63\,373--63\,394, 2019.

\bibitem{9154315}
A.~Zaidi and I.~E. Aguerri, ``Distributed deep variational information
  bottleneck,'' in \emph{2020 IEEE 21st International Workshop on Signal
  Processing Advances in Wireless Communications (SPAWC)}, 2020, pp. 1--5.

\bibitem{e22020213}
\BIBentryALTinterwordspacing
Y.~Uğur, G.~Arvanitakis, and A.~Zaidi, ``Variational information bottleneck
  for unsupervised clustering: Deep {G}aussian mixture embedding,''
  \emph{Entropy}, vol.~22, no.~2, 2020. [Online]. Available:
  \url{https://www.mdpi.com/1099-4300/22/2/213}
\BIBentrySTDinterwordspacing

\bibitem{9965818}
T.-H. Huang, A.~E. Gamal, and H.~El~Gamal, ``On the multi-view information
  bottleneck representation,'' in \emph{2022 IEEE Information Theory Workshop
  (ITW)}, 2022, pp. 37--42.

\bibitem{8986754}
Y.~Uğur, I.~E. Aguerri, and A.~Zaidi, ``Vector {G}aussian {CEO} problem under
  logarithmic loss and applications,'' \emph{IEEE Transactions on Information
  Theory}, vol.~66, no.~7, pp. 4183--4202, 2020.

\bibitem{wang2019deep}
Q.~Wang, C.~Boudreau, Q.~Luo, P.-N. Tan, and J.~Zhou, ``Deep multi-view
  information bottleneck,'' in \emph{Proceedings of the 2019 SIAM International
  Conference on Data Mining}.\hskip 1em plus 0.5em minus 0.4em\relax SIAM,
  2019, pp. 37--45.

\bibitem{Xu2014LMMVIB}
C.~Xu, D.~Tao, and C.~Xu, ``Large-margin multi-view information bottleneck,''
  \emph{IEEE Transactions on Pattern Analysis and Machine Intelligence},
  vol.~36, no.~8, pp. 1559--1572, 2014.

\bibitem{wan2021multi}
Z.~Wan, C.~Zhang, P.~Zhu, and Q.~Hu, ``Multi-view information-bottleneck
  representation learning,'' in \emph{Proceedings of the AAAI Conference on
  Artificial Intelligence}, vol.~35, no.~11, 2021, pp. 10\,085--10\,092.

\bibitem{wyner1975common}
A.~Wyner, ``The common information of two dependent random variables,''
  \emph{IEEE Transactions on Information Theory}, vol.~21, no.~2, pp. 163--179,
  1975.

\bibitem{disGrad21}
E.~Sula and M.~C. Gastpar, ``Common information components analysis,''
  \emph{Entropy}, vol.~23, no.~2, 2021.

\bibitem{5766249}
G.~Xu, W.~Liu, and B.~Chen, ``Wyners common information for continuous random
  variables - a lossy source coding interpretation,'' in \emph{2011 45th Annual
  Conference on Information Sciences and Systems}, 2011, pp. 1--6.

\bibitem{wynerCont16}
------, ``A lossy source coding interpretation of {W}yner’s common
  information,'' \emph{IEEE Transactions on Information Theory}, vol.~62,
  no.~2, pp. 754--768, 2016.

\bibitem{relaxWynerInfo22}
E.~Sula and M.~Gastpar, ``The {G}ray-{W}yner network and {W}yner’s common
  information for {G}aussian sources,'' \emph{IEEE Transactions on Information
  Theory}, vol.~68, no.~2, pp. 1369--1384, 2022.

\bibitem{kumar2014exact}
G.~R. Kumar, C.~T. Li, and A.~El~Gamal, ``Exact common information,'' in
  \emph{2014 IEEE International Symposium on Information Theory}.\hskip 1em
  plus 0.5em minus 0.4em\relax IEEE, 2014, pp. 161--165.

\bibitem{cover1999elements}
T.~M. Cover, \emph{Elements of information theory}.\hskip 1em plus 0.5em minus
  0.4em\relax John Wiley \& Sons, 1999.

\bibitem{donoho2000high}
D.~L. Donoho \emph{et~al.}, ``High-dimensional data analysis: The curses and
  blessings of dimensionality,'' \emph{AMS math challenges lecture}, vol.~1,
  no. 2000, p.~32, 2000.

\bibitem{10.1093/imamat/24.1.59}
\BIBentryALTinterwordspacing
R.~B. Marimont and M.~B. Shapiro, ``{Nearest Neighbour Searches and the Curse
  of Dimensionality},'' \emph{IMA Journal of Applied Mathematics}, vol.~24,
  no.~1, pp. 59--70, 08 1979. [Online]. Available:
  \url{https://doi.org/10.1093/imamat/24.1.59}
\BIBentrySTDinterwordspacing

\bibitem{4766926}
G.~V. Trunk, ``A problem of dimensionality: A simple example,'' \emph{IEEE
  Transactions on Pattern Analysis and Machine Intelligence}, vol. PAMI-1,
  no.~3, pp. 306--307, 1979.

\bibitem{kingma2013auto}
D.~P. Kingma and M.~Welling, ``Auto-encoding variational {B}ayes,'' \emph{arXiv
  preprint arXiv:1312.6114}, 2013.

\bibitem{poole2019variational}
B.~Poole, S.~Ozair, A.~Van Den~Oord, A.~Alemi, and G.~Tucker, ``On variational
  bounds of mutual information,'' in \emph{International Conference on Machine
  Learning}.\hskip 1em plus 0.5em minus 0.4em\relax PMLR, 2019, pp. 5171--5180.

\bibitem{8588399}
C.~Zhang, J.~Bütepage, H.~Kjellström, and S.~Mandt, ``Advances in variational
  inference,'' \emph{IEEE Transactions on Pattern Analysis and Machine
  Intelligence}, vol.~41, no.~8, pp. 2008--2026, 2019.

\bibitem{blei2017variational}
D.~M. Blei, A.~Kucukelbir, and J.~D. McAuliffe, ``Variational inference: A
  review for statisticians,'' \emph{Journal of the American statistical
  Association}, vol. 112, no. 518, pp. 859--877, 2017.

\bibitem{8957256}
K.-L. Lim, X.~Jiang, and C.~Yi, ``Deep clustering with variational
  autoencoder,'' \emph{IEEE Signal Processing Letters}, vol.~27, pp. 231--235,
  2020.

\bibitem{kingma2014semi}
D.~P. Kingma, S.~Mohamed, D.~Jimenez~Rezende, and M.~Welling, ``Semi-supervised
  learning with deep generative models,'' \emph{Advances in neural information
  processing systems}, vol.~27, 2014.

\bibitem{farnia2016minimax}
F.~Farnia and D.~Tse, ``A minimax approach to supervised learning,''
  \emph{Advances in Neural Information Processing Systems}, vol.~29, 2016.

\bibitem{alemi2016deep}
A.~A. Alemi, I.~Fischer, J.~V. Dillon, and K.~Murphy, ``Deep variational
  information bottleneck,'' \emph{arXiv preprint arXiv:1612.00410}, 2016.

\bibitem{720245}
J.~Treichler, M.~Larimore, and J.~Harp, ``Practical blind demodulators for
  high-order {QAM} signals,'' \emph{Proceedings of the IEEE}, vol.~86, no.~10,
  pp. 1907--1926, 1998.

\bibitem{vilaplana2014queuing}
J.~Vilaplana, F.~Solsona, I.~Teixid{\'o}, J.~Mateo, F.~Abella, and J.~Rius, ``A
  queuing theory model for cloud computing,'' \emph{The Journal of
  Supercomputing}, vol.~69, pp. 492--507, 2014.

\bibitem{anastasi2000qos}
G.~Anastasi and L.~Lenzini, ``{QoS} provided by the {IEEE} 802.11 wireless
  {LAN} to advanced data applications: a simulation analysis,'' \emph{Wireless
  Networks}, vol.~6, pp. 99--100, 2000.

\bibitem{kay1993fundamentals}
S.~M. Kay, \emph{Fundamentals of statistical signal processing: estimation
  theory}.\hskip 1em plus 0.5em minus 0.4em\relax Prentice-Hall, Inc., 1993.

\bibitem{li2022traffic}
Y.~Li, Y.~Huang, S.~Seneviratne, K.~Thilakarathna, A.~Cheng, G.~Jourjon,
  D.~Webb, D.~B. Smith, and R.~Y. Da~Xu, ``From traffic classes to content: A
  hierarchical approach for encrypted traffic classification,'' \emph{Computer
  Networks}, vol. 212, p. 109017, 2022.

\bibitem{xu2015empirical}
B.~Xu, N.~Wang, T.~Chen, and M.~Li, ``Empirical evaluation of rectified
  activations in convolutional network,'' \emph{arXiv preprint
  arXiv:1505.00853}, 2015.

\bibitem{srivastava2014dropout}
N.~Srivastava, G.~Hinton, A.~Krizhevsky, I.~Sutskever, and R.~Salakhutdinov,
  ``Dropout: a simple way to prevent neural networks from overfitting,''
  \emph{The journal of machine learning research}, vol.~15, no.~1, pp.
  1929--1958, 2014.

\bibitem{kingma2014adam}
D.~P. Kingma and J.~Ba, ``{ADAM}: A method for stochastic optimization,''
  \emph{arXiv preprint arXiv:1412.6980}, 2014.

\bibitem{xie2016unsupervised}
J.~Xie, R.~Girshick, and A.~Farhadi, ``Unsupervised deep embedding for
  clustering analysis,'' in \emph{International conference on machine
  learning}.\hskip 1em plus 0.5em minus 0.4em\relax PMLR, 2016, pp. 478--487.

\bibitem{kirchler2016tracked}
M.~Kirchler, D.~Herrmann, J.~Lindemann, and M.~Kloft, ``Tracked without a
  trace: linking sessions of users by unsupervised learning of patterns in
  their dns traffic,'' in \emph{Proceedings of the 2016 ACM Workshop on
  Artificial Intelligence and Security}, 2016, pp. 23--34.

\bibitem{scikit-learn}
F.~Pedregosa, G.~Varoquaux, A.~Gramfort, V.~Michel, B.~Thirion, O.~Grisel,
  M.~Blondel, P.~Prettenhofer, R.~Weiss, V.~Dubourg, J.~Vanderplas, A.~Passos,
  D.~Cournapeau, M.~Brucher, M.~Perrot, and E.~Duchesnay, ``Scikit-learn:
  Machine learning in {P}ython,'' \emph{Journal of Machine Learning Research},
  vol.~12, pp. 2825--2830, 2011.

\bibitem{9671851}
G.~Ke, Z.~Hong, Z.~Zeng, Z.~Liu, Y.~Sun, and Y.~Xie, ``Conan: Contrastive
  fusion networks for multi-view clustering,'' in \emph{2021 IEEE International
  Conference on Big Data (Big Data)}, 2021, pp. 653--660.

\bibitem{tian2020contrastive}
Y.~Tian, D.~Krishnan, and P.~Isola, ``Contrastive multiview coding,'' in
  \emph{European conference on computer vision}.\hskip 1em plus 0.5em minus
  0.4em\relax Springer, 2020, pp. 776--794.

\end{thebibliography}


\end{document}